# Momentum microscopy of ultrafast electron emission from a strongly driven optical nanoantenna

K. Harland[1,†], G. Hergert[1,†,*], Z. Pápa[2], X. Wu[3], L. Hansen[1], J. Altenburg[1], K. Meier[1], A. Klösgen[1], B. Hecht[4], J.-S. Huang[3,5], P. Dombi[6], J. Vogelsang[1,*]

[1]Institut für Physik, Carl von Ossietzky Universität Oldenburg, Oldenburg, Germany

[2]ELI ALPS, The Extreme Light Infrastructure ERIC, Szeged, Hungary

[3]Leibniz Institute of Photonic Technology, Jena, Germany

[4]Physikalisches Institut, Universität Würzburg, Würzburg, Germany

[5]Institute of Physical Chemistry and Abbe Center of Photonics, Friedrich-Schiller-Universität Jena, Jena, Germany

[6]HUN-REN Wigner Research Center for Physics, Budapest, Hungary

* Corresponding authors: germann.hergert@uol.de, jan.vogelsang@uol.de

† These authors contributed equally

## Abstract

Metallic nanostructures in combination with femtosecond lasers are a well-suited platform for the control of photoelectrons by strong and nano-localized driving fields, with high relevance for ultrafast, coherent electron emitters and petahertz electronics. Photoelectron spectroscopy resolves such photoelectron dynamics, but lacks nanoscale spatial resolution, making it only suited for single emitters or homogeneous arrays. We report the first strong-field experiment combining both photoemission electron microscopy and momentum microscopy, two complementary techniques providing spatial and momentum resolution within the same instrument. We apply this new methodology to a double-hole nanoantenna with sub-10 nm apex radii, demonstrating the potential of this approach for the control of photoelectrons in heterogeneous nanostructured samples using few-cycle light fields. Our measurements reveal distinct signatures of two classes of electron trajectories in the near-field. Quiver trajectories result in directed, angularly more focused emission, whereas subcycle trajectories give rise to a broader transverse momentum distribution. Surprisingly, this observation disagrees with previously reported emission characteristics from nanotip emitters, which we classify as a special case of a broader class of curved emitter surfaces driven by ultrashort light fields. This demonstrates both the impact of sophisticated electron detection methods and the potential of strong-field control of electrons in nanoscale geometries.

## 1. Introduction

Angle-resolved photoemission spectroscopy (ARPES) and momentum microscopy (MM) provide direct access to the occupied electronic band structure of solids through linear photoemission. The absorption of a high-energy photon promotes an electron into the vacuum while conserving its lateral momentum in the solid, thereby providing direct access to the occupied electronic states. When an additional strong optical field is applied, electrons inside the material can become transiently dressed by the driving field, leading to a modification of the band structure. This light-induced control has established the field of Floquet engineering [1-4], where electronic properties are manipulated on femtosecond timescales.

In contrast, in the regime of strong-field photoemission, the final photoelectron spectrum is no longer primarily determined by the initial band structure, but rather governed by the electron dynamics in the few-cycle laser field after emission [5-7]. In particular, a periodic quiver motion in the oscillating light field and rescattering at the material-vacuum interface strongly influence the observed energy and momentum distributions [8]. For nanostructures, these effects are further enriched by the presence of highly localized optical near-fields, which exhibit strong spatial gradients and rapid decay on the nanometer scale. These near-fields can significantly influence electron trajectories and offer a route to control photoelectrons on the nanoscale [9, 10].

Photoemission electron microscopy (PEEM) and MM provide two complementary approaches to imaging photoemitted electrons from surfaces, yielding real-space and in-plane momentum information, respectively. Although the two techniques are often regarded as distinct methods, both are based on the same electron-optical column, where the operation mode is selected by adjusting the excitation of the electrostatic electron optics to project either the emission position (PEEM) or the in-plane momentum (MM) onto the detector (see Extended Data Figure 2). Modern instruments therefore allow straightforward switching between real-space and momentum-space imaging. The information is supplemented by the simultaneous energy-resolved detection of photoelectrons using time-of-flight analysis. Combined with visible or near-infrared few-cycle laser pulses, PEEM and MM operate in a multiphoton or strong-field emission regime. Due to the nonlinear dependence of the photoemission yield on the local field strength, emission is strongly enhanced at regions with strong optical near-fields that stem from modal interference [11, 12], optical resonances [13], nanoscale features [14-16] or plasmonic modes [17, 18]. The latter results in highly confined emission regions, typically at sharp geometrical points such as nano-tips [19-22], antenna apices [23] or other field hot spots that might stem from plasmon or disorder induced localization [24-26].

The ability to alternate between real-space and momentum-space imaging within the same instrument enables powerful approaches that combine spatial and momentum selectivity. A recent example is dark-field imaging of excitons by Schmitt et al. [27], where a momentum aperture placed in an intermediate image plane selectively transmits photoelectrons originating from excitonic states. The resulting real-space images therefore directly map exciton dynamics with high spatial resolution. Vice versa, spatial apertures can be used to restrict momentum analysis to selected regions of a sample for investigating spatially heterogeneous systems.

Strong-field photoemission has so far been investigated primarily by photoelectron spectroscopy using either single nanoemitters or homogeneous emitter arrays [28, 29]. Long-wavelength excitation enables access to the strong-field regime at comparatively low incident intensities [30], resulting in localized electron emission with reduced space-charge effects that could influence spectral features [31]. However, conventional photoelectron spectroscopy averages the emitted signal over the illuminated region and therefore lacks the spatial resolution required to investigate individual nanoemitters or heterogeneous nanophotonic structures. Extending strong-field photoemission to PEEM and momentum microscopy therefore provides an opportunity to directly correlate localized electron emission with the underlying nanostructure and its associated momentum distributions.

Here, we combine strong-field photoemission with PEEM and MM to investigate the ultrafast electron dynamics of individual nanoemitters with simultaneous real-space and momentum-space resolution. We study a double-nanohole antenna structure featuring two opposing nanoscale apices and drive nonlinear photoemission using few-cycle short-wavelength infrared (SWIR) laser pulses. Real-space imaging reveals localized emission from both antenna apices, while momentum-resolved measurements provide insight into the in-plane electron acceleration. Primarily, we observe electrons that are accelerated in apex direction, with an additional weaker contribution of electrons accelerated perpendicularly. By comparing the measured distributions with simulations based on the simple man model and realistic emitter geometries, we identify distinct contributions from quiver and subcycle electrons. While oscillating quiver electrons remain in the enhanced near-field for multiple half-cycles of the light field, subcycle electrons are directly ejected from it. In contrast to previous observations in nanotaper systems, where subcycle electrons exhibited directed emission [10, 32], we find that subcycle electrons from realistic antenna geometries are emitted over a broad angular range and quiver electrons are preferentially accelerated in apex direction. We attribute this difference to the vectorial structure of the near-field, in particular the tangential field component arising from the local curvature of the emitter surface. Our results demonstrate that nanoscale geometry and near-field topology play an important role in shaping strong-field electron emission by few-cycle light fields and establish momentum microscopy as a tool for resolving strong-field dynamics from individual nanostructures.

## 2. Results and discussion

Strong-field studies on the double-hole antenna are performed in a PEEM/MM (Focus GmbH) at a pressure of $< 5 \times 10^{-9}$ mbar, in combination with SWIR pulses from a home-built amplifier system. The antenna structure is produced by helium ion beam milling (ZEISS Orion NanoFab) completely through a 40-nm thick monocrystalline gold (Au) flake [33] that is subsequently transferred onto a 500-μm thick silicon (Si) substrate. The substrate is p-type doped with Boron at a density of $5 \times 10^{18}\ \mathrm{cm}^{-3}$ to reduce its ohmic resistance and provide sufficient electrical conductivity of the sample. To drive nonlinear photoemission from the antenna, laser pulses are generated in a home-built amplifier system operating at a high repetition rate of 200 kHz and providing 20-fs pulses at a wavelength of 2000 nm with a passively stabilized carrier-envelope phase [34]. Illumination of the antenna is performed from the side opposite the detector through the substrate (see Figure 1a) by focusing the laser using an off-axis parabolic mirror with an effective focal length of 33 mm. The polarization of the few-cycle pulses is aligned parallel to the antenna apices (in $x$-direction), generating a strongly enhanced optical near-field in the apex region (see Extended Data Figure 4). The photoemitted electrons are collected by the PEEM objective lens and the subsequent electron lenses are used to either image the real space or momentum space onto the detector plane, allowing us to resolve the emission site or the momentum distribution of the detected electrons with a resolution of 40 nm or 6 m$\text{Å}^{-1}$, respectively. For electron detection, we use a microchannel plate followed by a luminescent phosphor screen and a complementary metal-oxide-semiconductor camera to resolve the two-dimensional electron distribution. Figure 1b shows a recorded scanning electron microscope (SEM) image used to characterize the antenna, revealing apex radii of approximately 10 nm, spaced by 350 nm. Photoemission triggered by the SWIR pulses and recorded in the PEEM in real-space mode is seen as the colored map in Figure 1c, showing that the emission is localized to the apex regions. Stronger emission is observed at the left apex compared to the right one, potentially caused by small, inhomogeneous surface contaminations. This asymmetry in the photoemission yield is further enhanced by the high nonlinearity of the emission process. An underlying gray image shows the antenna structure in the PEEM, obtained by blocking the SWIR laser and using ultraviolet (UV) light from a mercury-vapor gas-discharge lamp to trigger linear photoemission from the sample surface.

The strong optical near-field at the antenna apices accelerates the emitted electrons on sub-optical-cycle timescales, giving rise to a range of different electron trajectories depending on the instantaneous near-field amplitude and emission phase, as illustrated schematically in Figure 1d (light to dark purple lines). At lower field strengths the electron motion shows a periodic quiver motion or rescattering in the oscillating near-field, while increasing field amplitudes result in an acceleration of the electrons out of the near-field within less than an optical cycle, giving rise to subcycle electron motion.

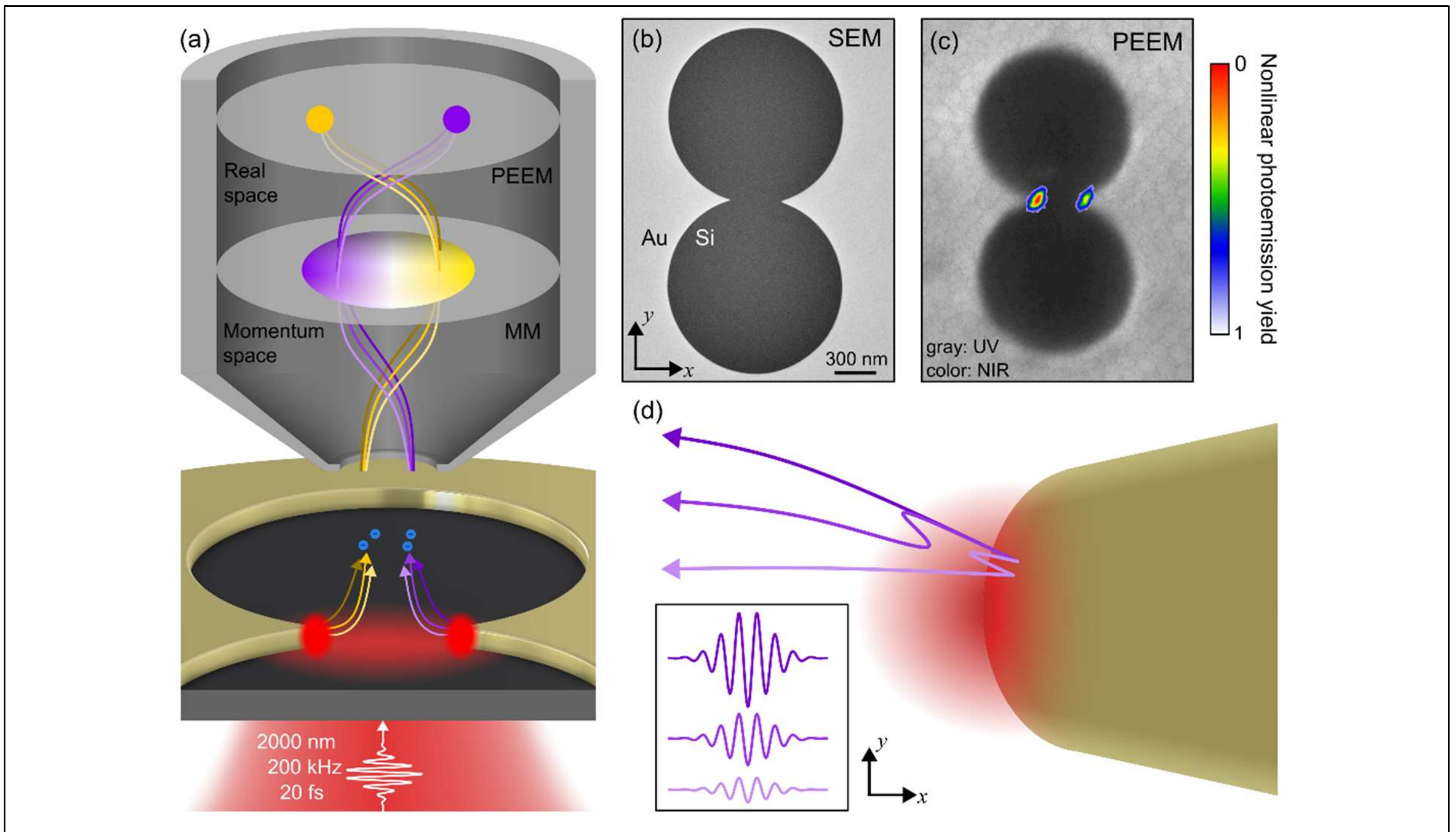

**Figure 1:** Experimental setup for real-space and momentum-space imaging of photoemission from a double-hole nanoantenna in a photoemission electron microscope (PEEM). **(a)** An intense short-wavelength infrared (SWIR) pulse with a pulse duration of 20 fs is focused through a silicon substrate onto a double-hole nanoantenna that is milled into a monocrystalline gold flake placed on the substrate. The high field enhancement at the apices of the antenna triggers locally restricted strong-field photoemission from the left and right apex (drawn in yellow and purple, respectively) that is collected by the objective lens of the PEEM. The lenses in the PEEM column allow to switch between momentum- or position-resolved imaging of the photoemission at the detector plane. **(b)** Scanning electron microscope (SEM) image of the antenna. The two opposing apices have a distance of 350 nm to each other and an apex radius below 10 nm. **(c)** Real-space image of the nonlinear photoemission (color) from the antenna apices triggered by the SWIR pulses measured in the PEEM. The overlay (gray) is acquired by linear photoemission using ultraviolet (UV) light from a mercury lamp. The faint hexagonal structure is an artifact of the imaging unit. **(d)** Schematic electron trajectories in the near-field from the optically excited antenna apex, with increasing near-field amplitude (light to dark purple line). Different electron dynamics are possible, namely rescattering, quiver and subcycle motion.

We begin by analyzing the in-plane $(k_x, k_y)$ photoelectron momentum distributions from the nanoantenna for various local near-field intensities $I_{\mathrm{NF}}$ between $12.0$ and $40.8\ \mathrm{TW/cm^2}$ (see Figure 2a-c). We estimate the local near-field intensity from the measured laser power and pulse duration in front of the sample by determining the focused SWIR spot size directly in the PEEM (see Extended Data Figure 3) and using a field enhancement factor of 20 for both apices that is obtained by simulating the near-field of a realistic sample geometry using finite-difference time-domain simulations (see Methods). At the lowest intensity (see Figure 2a) the observed momentum distribution is nearly circular in all directions. With increasing intensity, we primarily observe a broadening of the momentum distribution in positive and negative $k_x$-direction (see Figure 2b), i.e., in direction of the antenna apices, as a result of field-driven acceleration in the antenna near-fields. The electrons that are emitted from the left apex (see Figure 1b) are accelerated toward positive $k_x$ and vice versa. The acceleration and hence broadening of the distribution in positive $k_x$ is more pronounced as a consequence of the stronger near-field enhancement at the left apex that is apparent from the higher photoemission yield from this apex in Figure 1b. For comparison, we show the results from a similar antenna structure in Supporting information chapter 1, where only one apex shows significant photoemission and consequently the broadening in momentum space occurs in only one direction. With further increasing near-field intensity (see Figure 2c), the broadening in $k_x$ increases, while also a small broadening in $k_y$ becomes visible, which is most pronounced at the center $k_x = 0$.

By integrating the momentum distributions over the entire detector plane, we obtain the total number of detected electrons per laser pulse, which is measured for a larger range of near-field intensities (black dots, see Figure 2d). We observe the transition from multiphoton to strong-field photoemission indicated by the kink in the photoemission yield at around $25\,\mathrm{TW/cm^2}$ on the double-logarithmic axes. A fit, represented by the solid line, shows a nonlinear behavior of the photoemission yield of $\sim I_{NF}^{n}$, where $n = 5.6$ is the determined nonlinearity in the multiphoton photoemission regime. The difference to the expected nonlinearity given by the ratio of the gold work function ($\Phi_{\mathrm{Au}} = 4.8\,\mathrm{eV}$ [35]) and the photon energy at 2000 nm ($E_{\mathrm{ph}} = 0.62\,\mathrm{eV}$) of $n = \Phi_{\mathrm{Au}}/E_{\mathrm{ph}} \approx 7.7$, can be explained by atomically thin surface contaminations in combination with the broad laser spectrum permitting also photoemission channels with lower photon numbers. The red crosses in Figure 2d mark the intensities at which the shown momentum distributions have been recorded. Here, the distribution with the lowest intensity clearly is in the multiphoton regime, while the highest intensity reaches the strong-field regime with a total photoemission rate of ~65 electrons per pulse. During the determination of the count rates, the detection system is adjusted to ensure that every electron is counted. In the Supporting Information we discuss the minor influence of Coulomb repulsion effects, which we can safely ignore going forward.

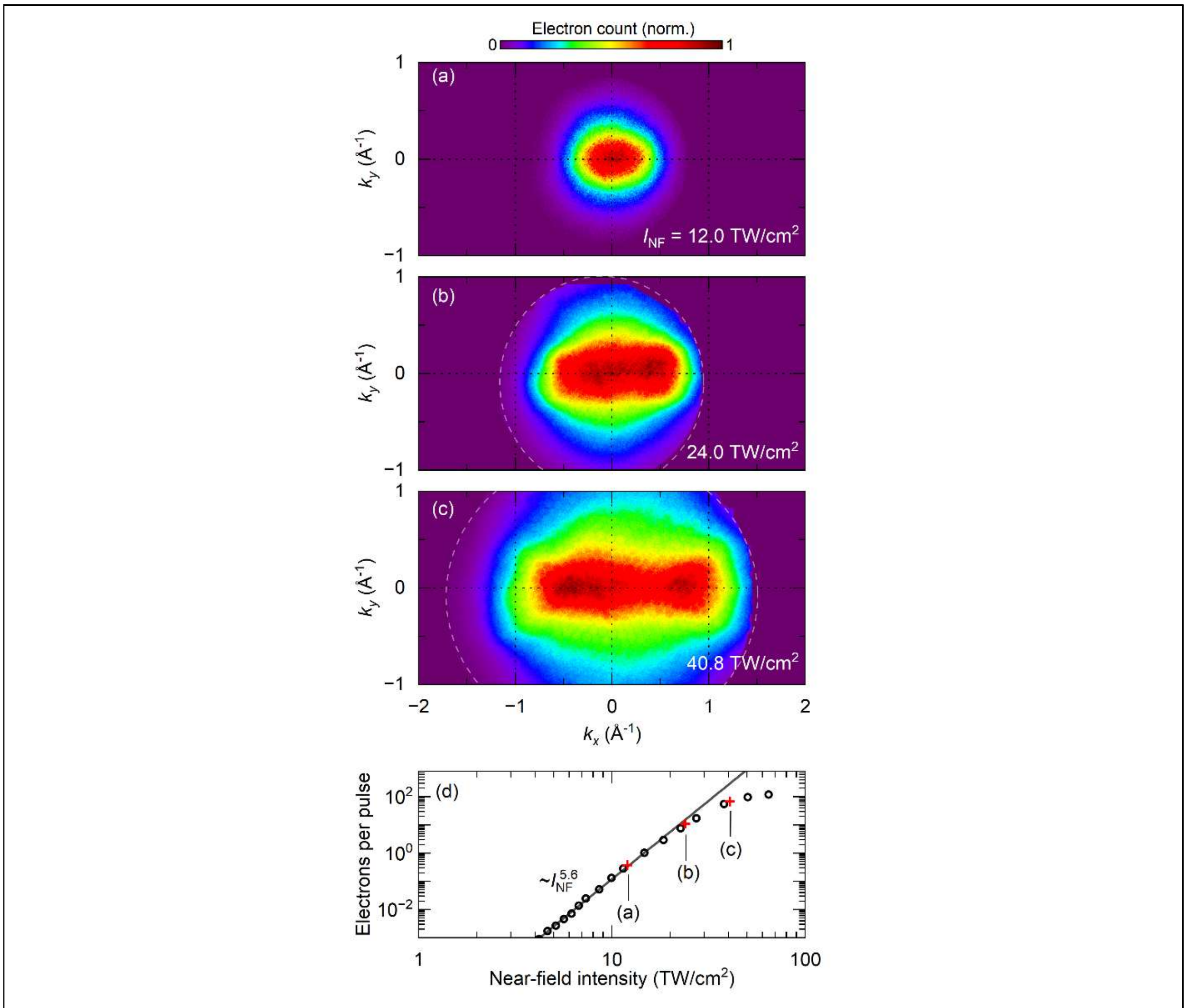

**Figure 2:** Momentum-resolved strong-field electron acceleration from a double-hole nanoantenna measured in a photoemission electron microscope. **(a-c)** In-plane $(k_x, k_y)$ momentum distributions (color-coded) of the photoelectrons for increasing near-field intensities $I_{\mathrm{NF}}$ driven by short-wavelength infrared pulses. For the lowest intensity (a) a narrow electron distribution is observed, which predominantly broadens in $k_x$-direction, i.e. in direction of the apices with increasing intensity as result of laser-driven electron acceleration (b, c). A broadening in $k_y$-direction is also observed and is most prominent at the center ($k_x = 0$) of the distributions. The dashed white lines mark the detector area. **(d)** Total photoelectron yield as a function of the local near-field intensity (circles). The yield shows a kink, indicating the transition into the strong-field emission regime. A photoemission nonlinearity of 5.6 is extracted by a linear fit in the low-intensity regime (line). The red crosses mark the intensities at which the data from (a-c) have been recorded.

We perform simulations based on classical electron motion in the optical near-field of a nanostructure that mimics the apex of the illuminated nanoantenna to analyze the resulting electron trajectories. Typically, ion-milled structures exhibit rounded edges as a consequence of the milling process with very high, but limited resolution [17, 36]. These edges are facing out of the sample plane and generate near-fields with electric field vectors oriented ~50° out of the plane, as shown by electromagnetic simulations [13]. We therefore model each apex as a nanospheroid with equal radii $r_x = r_z = 5\ \mathrm{nm}$ in the $x$-$z$-plane (along the apex direction and toward the PEEM column, see Figure 3a), corresponding to the typical feature size of helium-milled structures [17, 36], and a larger radius of $r_y = 10\ \mathrm{nm}$ perpendicular to the apex direction given by the experimentally determined apex radius in this direction (see Figure 3b). Electrons are numerically placed at different positions and emission times in the near-field of the nanospheroid, that is excited by a few-cycle laser pulse polarized at 50° in the $x$-$z$-plane and their trajectories are propagated independently of each other. We refer the interested reader to the Supporting Information for a discussion of why an independent propagation of electrons is justified here. From the three-dimensional final electron momenta, we extract the in-plane momentum distribution shown in Figure 3c, for a near-field intensity of $40\ \mathrm{TW/cm^2}$. To account for the opposing apex after simulating the emission from only one, we have added a mirrored image of the distribution along the $k_y$-axis. We achieve good agreement between the simulated and measured distributions. The elongated shape of the momentum distribution in $k_x$-direction is mainly a consequence of the prolate shape of the nanospheroid ($r_{x,z} < r_y$). When such a spheroid is excited along its short major axis ($x$), the generated near-field component $E_x$ is more pronounced than $E_y$ resulting in a primarily acceleration of the electrons in apex direction.

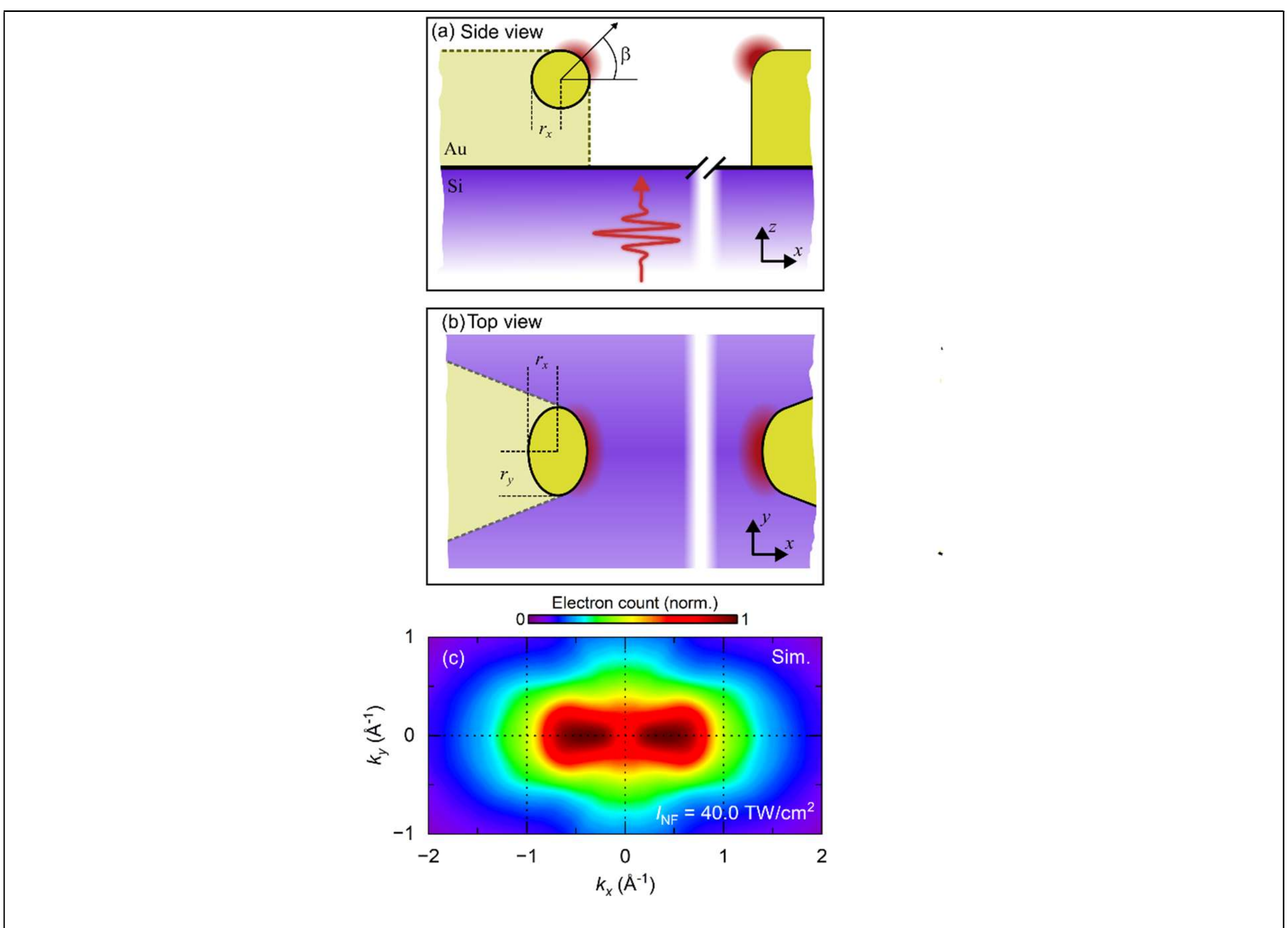

**Figure 3:** Numerical model for the near-field dynamics around the nanostructure. **(a, b)** Sketch of the side (a) and top (b) view of the nanospheroid model applied to the nanoantenna. **(c)** Simulated momentum distribution based on field-driven electron trajectories mimicking our experimentally measured distributions.

In order to analyze the in-plane angle characteristics of the photoemission process, we convert the final momentum distribution from the experiment and the simulation to final emission angles $\alpha = \tan^{-1}(k_y/k_x)$. For this, we integrate over all pixel values within one angular bin, resulting in a projection of the two-dimensional distributions onto a radial emission direction. The final angle distribution for five intensities, increasing from light to dark red lines, is shown in Figure 4e. For the lowest intensity, we observe a nearly angular-independent distribution. With increasing intensity, the distribution broadens toward 0° and 180° resulting from forward accelerated electrons from each respective apex. Clear features at 90° and 270° can be observed with increasing intensity, indicating an increasing contribution of photoelectrons that are emitted sideways in $y$-direction, i.e., perpendicular to the apex direction. For the highest intensity of $40.8\,\mathrm{TW/cm^2}$, the expected electron quiver amplitude is $l_\mathrm{q} = eE_\mathrm{NF}/(m\omega^2) = 3.5\,\mathrm{nm}$, where $\omega$ is the angular frequency of driving laser wavelength, $e$ is the elementary charge, $m$ the eletron's mass and $E_\mathrm{NF} = \sqrt{2I_\mathrm{NF}/c_0\varepsilon_0}$ is the near-field amplitude with the near-field intensity $I_\mathrm{NF}$, vacuum speed of light $c_0$ and the vacuum permittivity $\varepsilon_0$. The expected quiver amplitude is comparable to the apex dimension and hence also to the near-field localization length. Therefore, we can expect an increasing number of electrons that are photoemitted and accelerated out of the near-field within less than one optical half-cycle at the highest intensities. These subcycle electrons have been investigated in ultrafast photoemission from metallic nanotapers, where a predominantly forward acceleration of subcycle electrons has been reported resulting in a narrow emission cone [10, 32]. Here, however, we observe an increase in sideways accelerated subcycle electrons, which we also find in the simulated data shown in Figure 4e (blue line). This is a consequence of the electron acceleration in the near-field of our nanostructure. We will later find that small differences in the curvature of the surface lead to focusing of either subcycle or quiver trajectories, depending on the sign of a small tangential field component.

The momentum distributions are further analyzed in Figure 4a-d. We extract the width of the energy distribution from the measured and simulated in-plane momentum distributions, given by the kinetic energies corresponding to the two half-maximum positions in both $k_x$-directions (see Methods). For the experimental data, the two emission apices are analyzed separately and the widths of the energy distribution for emission in positive (red dots, Figure 4a) and negative (orange squares) $k_x$-direction are shown. The simulated intensity dependence of the width of the energy distribution is presented in Figure 4b. At the highest experimental near-field intensity ($40.8\,\mathrm{TW/cm^2}$), the pondermotive energy ($U_\mathrm{P} = e^2E_{NF}^2/(4m\omega^2)$) is $15.2\,\mathrm{eV}$ and the observed width of the energy distribution amounts to $\sim 0.7U_\mathrm{P}$ and $\sim 0.4U_\mathrm{P}$ in positive and negative $k_x$-direction, respectively. In the numerical simulation, the width of the energy distribution amounts to $\sim 0.4U_P$ as well. We note that we use the width of the energy distribution and not its cutoff to compare the measured and simulated distributions, because this work is focused on the low energy angular distribution of the photoelectrons and thus does not perfectly reproduce the behavior around the cutoff region.

This observed widths of the energy distribution are well below typical $10U_\mathrm{P}$ cutoff values that hold for spatially extended optical near-fields [37] or for photoionization from atoms [38] and molecules [39], where the quiver motion and subsequent rescattering dynamics of the electrons impact the kinetic energy spectra. At our present conditions, where the electron quiver amplitude is $l_\mathrm{q} = 3.5\ \mathrm{nm}$ for the highest applied intensities and the near-field decay length in the simulation is $l_\mathrm{NF} = 3.4\ \mathrm{nm}$, quiver motion is substantially suppressed: The parameters place the emission at the transition to the subcycle regime, characterized by the adiabaticity parameter $\delta = l_\mathrm{NF}/l_\mathrm{q} \approx 1$ [9, 40]. In this regime, the electron no longer oscillates in the driving near-field, and thus, the ponderomotive energy is no longer the relevant energy scale for the final photoelectron spectrum. Instead, the electron leaves the spatiotemporally localized near-field within less than one optical half-cycle, preventing the field from transferring the full ponderomotive energy to the electron. This characterizes the transition to the subcycle regime and explains the observed width of the energy distribution well below the cutoff value of $10U_\mathrm{P}$. As the near-field intensity increases further, the electron spends even less time in the near-field, reducing the efficiency of momentum and energy transfer. Consequently, the width of the energy distribution exhibits a sub-linear scaling with the near-field intensity [41]. In fact, this behavior is clearly reproduced in the simulations (Figure 4b), and can be observed as well in the $-k_x$ direction in the experimental data (Figure 4a). In the $+k_x$ direction, a quite pronounced redistribution of momenta for different near-field intensities prevents a direct comparison with the simulations. A cross section through the center of the distribution at $k_x = 0$ along $k_y$ for increasing near-field intensity is presented for the experiment and simulation in Figure 4c and d, respectively. For the lower intensities, we observe Gaussian-shaped profiles with increasing widths. Beginning at $24\ \mathrm{TW/cm^2}$, in the experiment, clear shoulders arise around $k_y = \pm 1 \mathrm{Å}^{-1}$. These shoulders are comprised of the perpendicular accelerated electrons that are also visible in the polar plot at 90° and 270° (Figure 4e) and become particularly relevant in the strong-field regime, where the confined optical near-field increases the relative amount of subcycle electrons.

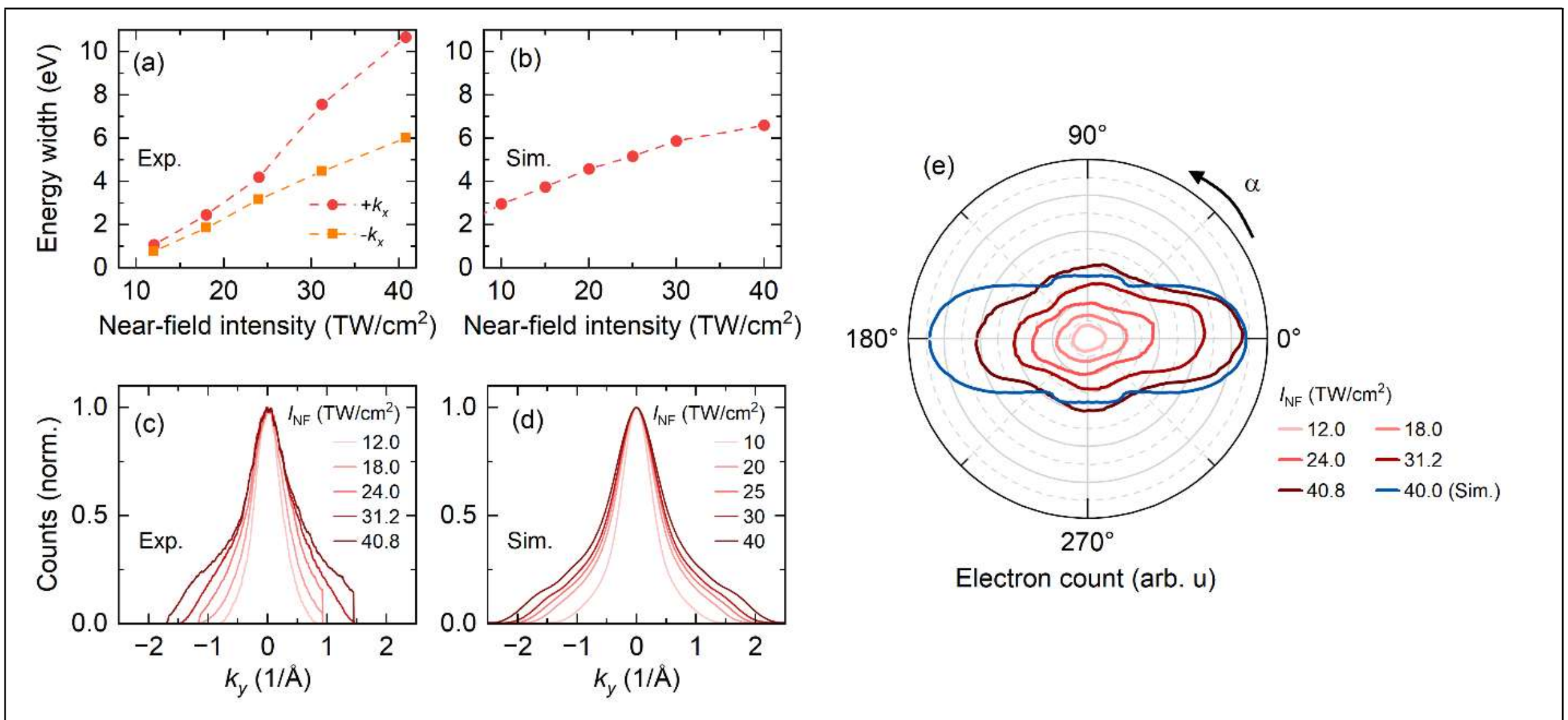

**Figure 4:** Analysis of experimental and simulated momentum-resolved photoemission from the nanoantenna. **(a, b)** Width of the energy distribution for experiment (a) and simulation (b). For the experiment, we show the width for both apices separately from which electrons are accelerated in positive (red dots) and negative (orange squares) $k_x$-direction. The width is at $\sim 0.7 U_\mathrm{P}$ (for $+k_x$) and $\sim 0.4 U_\mathrm{P}$ (for $-k_x$) in the experiment at the highest intensity of $40.8\ \mathrm{TW/cm^2}$, with $U_\mathrm{P}$ the pondermotive energy. In the simulation the width of the energy distribution amounts to $\sim 0.4 U_\mathrm{P}$ for the highest intensity. **(c, d)** Intensity dependent cross sections along $k_y$ at $k_x = 0$. With increasing intensity (light to dark red lines), the cross sections in the experimental (c) and simulated (d) data show a Gaussian-shaped profile that develops successively stronger shoulders around $k_y = \pm 1 \text{Å}^{-1}$. This reflects the increasing contribution of subcycle electrons that are accelerated sideways. **(e)** Emission angles extracted from the measured momentum distributions for increasing intensities (red lines), showing forward (0°, 180°) and sideways (90°, 270°) accelerated electrons. The angular distribution extracted from a simulation (blue line) shows a similar emission pattern.

Deviations from radial acceleration, which we observe at $k_x = 0$, can occur due to small transverse electric fields, when the overall momentum of the electron reaches small values. This happens for trajectories in which the local laser field decelerates photoelectrons to nearly zero momentum. To demonstrate this influence of the nanoscale surface, we choose two different geometries for the nanospheroid, visualized by the yellow areas in Figure 5a and b, one where the radius $r_x$ in direction of the laser polarization is larger and one smaller than the radius $r_y$, respectively. We set the radii to $r_x = 5\ \mathrm{nm} < r_y = 10\ \mathrm{nm}$ (Figure 5a), resulting in photoemission from a surface with rather low curvature in perpendicular direction, and $r_x = 5\ \mathrm{nm} > r_y =$ $2.5\ \mathrm{nm}$ (Figure 5b), to generate a pointy, nanotip-like geometry for the emitter. For both cases, we place an electron at $z = 0$ and choose the $x, y$ coordinate such that the initial photoemission angle $\alpha_{\mathrm{start}}$, which we define by the direction of the surface normal vector, equals 20° for both cases. For illustration purposes, we set the initial photoemission time to 0.4 fs after the maximum of the driving field and, for the rounded case shown in Figure 5a, set the near-field strength to $10.0\ \mathrm{V/nm}$ (red line) and $30.0\ \mathrm{V/nm}$ (blue line), resulting in a quiver and subcycle movement, respectively. Interestingly, the quiver trajectory bends inward, toward the apex during its motion back to the emitter, resulting in a smaller final emission angle compared to the subcycle trajectory. We emphasize that neither the starting time nor the starting position of an electron changes the observed bending of the trajectory as long as the trajectory remains of quiver or subcycle type. For the tip-like case shown in Figure 5b, the near-field strengths are set to $2.0\ \mathrm{V/nm}$ and $10.0\ \mathrm{V/nm}$, with the trajectories drawn in red and blue, respectively. Here, the subcycle trajectory exhibits the smaller emission angle, as also observed by others [10]. The inset shows a close-up of the motion near the surface, revealing the opposite bending, i.e. outwards, of the quiver trajectory.

The two different quiver trajectories for the two surface shapes can be understood by examining the optical near fields of the two geometries. Figures 5a and b additionally visualize the direction of the tangential component of the optical force acting on the electrons in the $x$-$z$-plane as gray arrows. Due to the curved surface combined with linearly polarized excitation, already in a small distance to the spheroid the field lines are no longer perfectly perpendicular to the surface. The resulting small tangential field component governs the electron motion parallel to the emitter surface after an electron has been emitted in an on average normal direction. The arrows are shown for the optical phase at which the electrons are driven back toward the emitter. During this half-cycle of the optical field, the electrons are decelerated and reach their lowest kinetic energy before being accelerated toward the surface. This is the critical time where the small tangential field component defines the tangential movement of the electron. In the apex region, the direction of the tangential force differs markedly between the two geometries. For the rounded geometry, the force is directed toward the center of the apex, whereas for the tip-like geometry it points away from the apex.

Consequently, the quiver trajectories are deflected in opposite directions, giving rise to the distinct angular photoemission characteristics observed in the experiments.

We repeat the simulations over a broader parameter space, varying both the near-field strength and the emitter radius $r_y$. The resulting final photoemission angles are shown as a color map in Figure 5c. The white lines separate regions corresponding to different classes of electron trajectories. In general, rescattered electrons are emitted at small angles (<20°), whereas quiver electrons emerge at slightly larger angles (~17°–25°). The comparatively small detection angles of rescattered electrons result from the assumed ballistic reflection at the surface, which redirects them toward the center of the apex. This finding contradicts the intuitive expectation that an electron emitted directly on the $x$-axis would experience a deflection in $y$-direction during rescattering when it does not hit the apex perfectly in the center. Conversely, directed forward emission exhibits insensitivity to minor lateral deviations in the trajectory of backscattered electrons, thereby ensuring the stability of this emission direction. In contrast, subcycle electrons exhibit the broadest angular distribution within the investigated parameter range. Their final emission angle increases with increasing emitter radius, indicating broader emission from rounded emitters than from more sharply curved ones. The largest emission angles occur in the limit of a nanosphere (grey dashed line), where the near field possesses the strongest field components in $y$-direction.

We further analyze the sensitivity of the final emission angle to the near-field strength by taking the derivative of the map in Figure 5c with respect to $E_{\mathrm{NF}}$, shown in Figure 5d. The respective sensitivity in the rescattering and quiver regime is rather complicated because of the exact timing the electron leaves the near-field and hence the number of optical cycles it interacts with. For the subcycle regime, however, a clear trend emerges. We observe that, with increasing near-field amplitude, the final emission angle for nanospheroids with $r_y > 3.5\ \mathrm{nm}$ increases gradually (blue color). In contrast, this trend reverses for $r_y < 3.5\ \mathrm{nm}$, where subcycle electrons are emitted at progressively smaller angles (red color). Interestingly, the increasing emission angle is present for rounded emitters, the nanosphere case and even partially for tip-like geometries and is hence the more general case. Thus, the case of more directed electron acceleration in the subcycle emission case, which has so far been solely discussed in the literature, is rather a special case of more complex dynamics with mostly the opposite behavior.

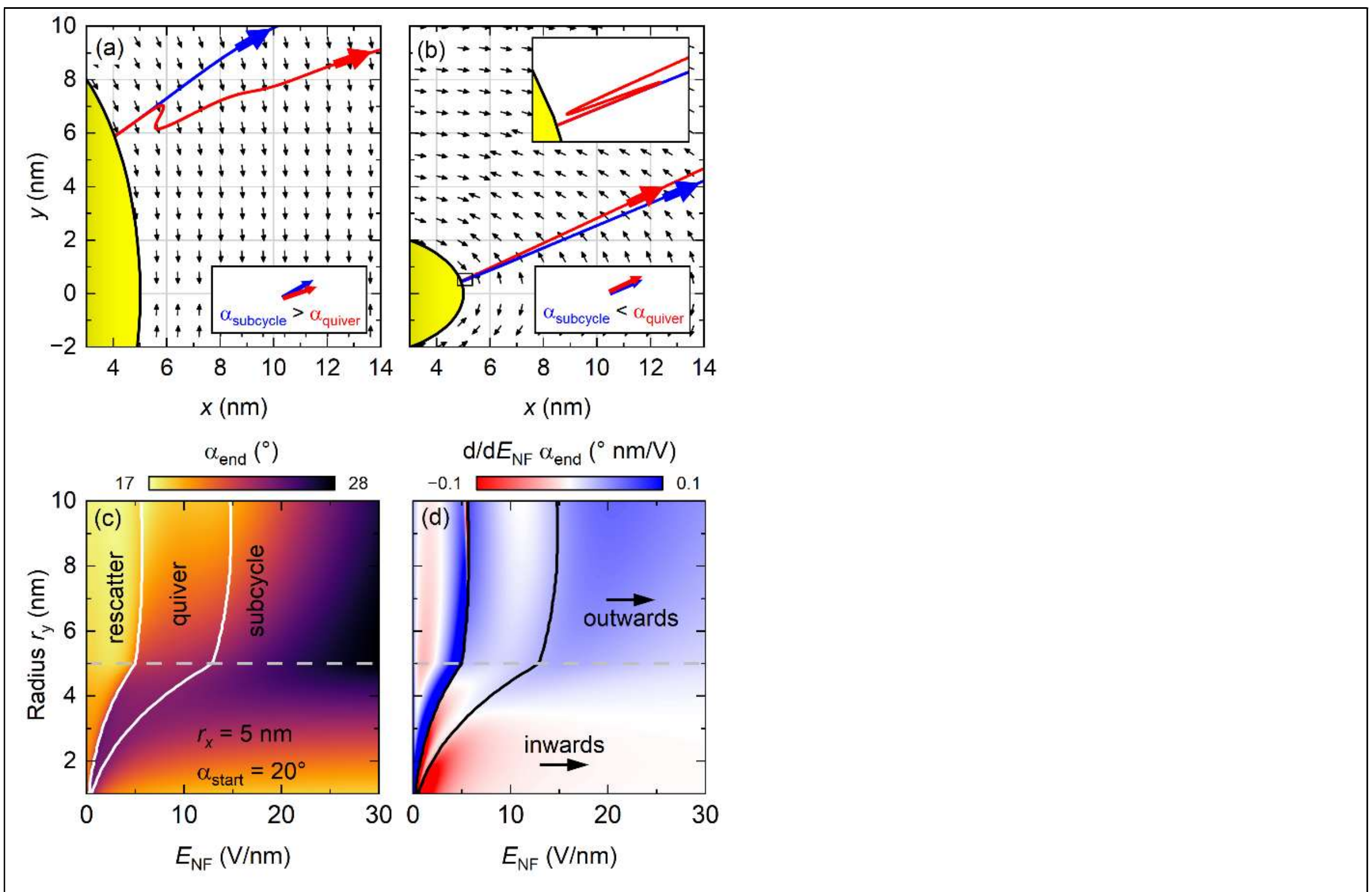

**Figure 5:** Electron trajectories for nanospheroids with different aspect ratios. (**a, b**) Representative electron trajectories for nanospheroids with $r_x = 5\,\mathrm{nm}$ and $r_y = 10\,\mathrm{nm}$ (a) or $r_y = 2.5\,\mathrm{nm}$ (b), corresponding to flat and tip-like geometries, respectively. Subcycle (blue) and quiver (red) trajectories are shown for an initial emission angle of $\alpha_{\mathrm{start}} = 20°$. The inset in (b) shows a zoom into the trajectories close to the surface. The electric force is almost perpendicular to the surface, but gray arrows indicate the direction of the very small tangential force component in the near-field when electrons are driven back toward the surface. The different field directions near the apex result in opposite deviations of the quiver trajectories for the two geometries. **(c)** Final emission angle $\alpha_{\mathrm{end}}$ as a function of near-field strength $E_{\mathrm{NF}}$ and nanospheroid aspect ratio. Solid lines indicate transitions between rescattering, quiver, and subcycle regimes. The dashed line marks the spherical case ($r_x = r_y$). **(d)** Sensitivity of the final emission angle to the near-field strength, $\mathrm{d}/\mathrm{d}E_{\mathrm{NF}}\,\alpha_{\mathrm{end}}$. While rescattered and quiver electrons show a complex dependence due to the exact timing of leaving the near-field, subcycle electrons exhibit a systematic trend. Here, an increasing near-field strength enhances outward deflection (blue) for flat geometries and inward deflection (red) for tip-like geometries.

## 3. Conclusion

In summary, we have demonstrated the first strong-field experiment performed in a photoemission electron microscope on a double-hole antenna driven by few-cycle laser pulses. Real-space images reveal localized nonlinear photoemission from the very apices of both antennas, while experimental in-plane momentum distributions reveal insights into the near-field driven electron trajectories and acceleration. For relatively weak driving fields, we observe a radial symmetric electron momentum distribution, while a clear directed emission pattern emerges in the strong-field regime. At intermediate field strengths, we primarily observe electron acceleration in apex direction. At the highest field strengths, an additional electron contribution in momentum space perpendicular to the apex direction is detected. Supported by simulations using the simple man model, we can assign the additional signal to subcycle electrons that leave the near-field of the emitter under larger angles compared to electrons that undergo a quiver motion. This trend is in contrast to experiments and simulations on tip-like structures that have been carried out in the past years. Upon detailed examination of the tangential near-field components, the behavior of directed subcycle electron emission observed to date turns out to be a special case of a limited parameter set. The opposite case, which is experimentally observed and analyzed here, occurs over a wider range of surface geometry parameters and focusses quiver electrons to a narrower emission angle compared to subcycle electrons. This behavior is independent of the emission place and time and thus also independent of the probability of certain trajectories relative to each other. A precise analysis and design of the tangential near-field component thus offers a new route to the coherent control of charge carriers around nanostructures using ultrashort laser pulses. The understanding of electron movement in optical near-fields is a fundamental requirement both for the design of ultrafast electron sources and the analysis of photoelectrons in strong-field driven, nanostructured samples. The latter case is an emerging field as optical field control of charge carriers inside a material like Floquet engineering becomes available in more and more (so far spatially homogeneous) material systems. The effect of the optical field on the charges in the material can only be understood when the subsequent motion of photoemitted electrons is also considered, thus revealing the intriguing dynamics inside the material before photoemission.

In a next step, we will utilize the time-of-flight capabilities of our device to record the fully three-dimensional momentum space information of photoelectrons, which is a critical milestone for the investigation of charge carrier dynamics in solids. An iris aperture in an intermediate real space plane will permit the analysis of individual emitters and regions of a sample in momentum space, driven by strong optical fields. This highlights the versatility of electron-optical imaging systems and illustrates how selective filtering in conjugate image planes provides a powerful route toward combined spatially- and momentum-resolved spectroscopy.

## SI Content

1) Momentum distribution from a double-hole antenna with one emitting apex
2) Estimations of Coulomb repulsion
3) Analytical near-fields of a prolate nanospheroid

## Author contributions

JV conceived the experiment. ZP and JV designed the sample with the help of FDTD simulations performed by ZP. XW fabricated the sample under the supervision of JSH in collaboration with BH. KM and AK prepared the optical setup and KH, JA, LH and GH performed the PEEM experiment under the supervision of JV. The experimental data was analyzed by GH, KH and JV and interpreted by GH, KH, PD and JV. Trajectory calculations were performed by GH supported by KH and JV. The first draft of the manuscript was written by GH and JV and all authors contributed to the discussion and final version of the manuscript.

## Funding

We acknowledge support from the ELI-ALPS project (GOP-1.1.1-12/B-2012-000 and GINOP-2.3.6-15-2015-00001), which is supported by the European Union and co-financed by the European Regional Development Fund. The work of Z.P. was supported by a Bolyai Research Scholarship of the Hungarian Academy of Sciences (MTA), project nr. BO/00773/24. J.-S.H. acknowledges support by the Deutsche Forschungsgemeinschaft DFG via SFB 1375 NOA (398816777; sub-project C1) and IRTG 2675 “Meta-Active” (437527638; sub-project C1). P.D. acknowledges support by the National Research, Development and Innovation Office of Hungary via project KKP137373. J.V. acknowledges support by the zukunft.niedersachsen program of the Niedersächsisches Ministerium für Wissenschaft und Kultur (DyNano and Stay Inspired) and the DFG (462448709, Emmy Noether program).

## Methods

### Generation of near-infrared, few-cycle pulses.

For the generation of few-cycle pulses in the short-wavelength infrared (SWIR) range used for the nonlinear photoemission, we utilize a sequence of nonlinear optical processes in a home-built setup (see Extended Data Figure 1). A detailed characterization of the laser setup has been recently published [34], while a compact description of the setup follows here.

The laser system is pumped by an ytterbium-doped fiber laser (Carbide from Light Conversion) with a repetition rate of 200 kHz, providing short laser pulses with a pulse duration of 200 fs at 1030 nm wavelength and an output power of 80 W. This fundamental beam is split into different fractions, used for pumping the following nonlinear processes. A first part of the pump laser is split off for supercontinuum generation in an yttrium-aluminum-garnet (YAG) crystal to generate a broad spectrum, extending down to a wavelength of 450 nm. The supercontinuum is temporally compressed using commercially available chirped mirrors and is afterwards focused into a beta-barium borate (BBO) crystal for noncollinear optical parametric amplification (NOPA) within a spectral range of 630 nm to 740 nm, pumped by the second harmonic of the fundamental input at 515 nm. Conversion to the short-wavelength infrared regime is achieved by difference frequency generation (DFG) in a BBO crystal, combining the NOPA pulses and a fraction of the fundamental pump, resulting in laser pulses centered around 2000 nm. In a final stage, the DFG pulses enter another BBO crystal for optical parametric amplification (OPA) pumped by a further part of the fundamental beam. Here, we reach an amplified DFG spectrum ranging from 1750 nm to 2400 nm with a total output power of 4.5 W, corresponding to 22.5 μJ pulse energy.

We use custom-made chirped mirrors and a pair of zinc sulfide (ZnS) wedges to compensate the second- and third-order spectral phase of the generated SWIR pulses for temporal post-compression. The compressed pulses are then either directed to the PEEM setup for the photoemission experiments or to a dispersion scan (d-scan) setup [42, 43] for characterization of the pulse duration and spectral phase. The beam path for the d-scan additionally includes replicas of the vacuum chamber window and the silicon substrate used in the experiment to enable an accurate comparison between the measured pulse duration and the conditions at the sample. For the d-scan measurement, the SWIR pulses are tightly focused onto the air–glass interface of a thin fused silica plate to generate third-harmonic (THG) radiation. The THG spectrum is recorded while the pulse dispersion is systematically varied by inserting additional ZnS from the wedge pair into the beam path. The resulting d-scan spectrogram is reconstructed using an open-source pulse retrieval algorithm (COPRA) [44], yielding the amount of wedge insertion for the shortest pulse duration and the flattest spectral phase, showing a pulse duration of 20 fs, measured by the full-width at half maximum (FWHM) of the intensity envelope.

**Nanoantenna fabrication.**

A monocrystalline gold flake grown on glass coverslips was selected and covered with a polymethyl methacrylate (PMMA) droplet. After 15-minute baking at 100 °C on a hot plate, the gold flake was stripped off from the glass coverslip with the PMMA droplet and transferred to another glass coverslip that was coated with a chromium (Cr) layer with void areas of different sizes. The gold flake was placed on a slightly smaller void area, such that the outer part of the gold flake was contacting the Cr layer (to avoid charging) but most of the area of the flake was contacting the glass. The PMMA droplet was dissolved in acetone after 10-minute baking at 120 °C on a hot plate. Outlines of the nanoantennas were then milled using a helium ion beam (HIM), on the area where the gold flake was contacting the glass. Using the same method, the patterned gold flake was stripped off from the glass coverslip with a PMMA droplet and transferred to a silicon substrate, while the inner parts of the outline of the nanoantennas were left on the glass coverslip. In the last step, the PMMA droplet was dissolved in acetone after 10-minute baking at 120 °C on a hot plate.

**Photoelectron setup.**

The nanoantenna, situated in the electron microscope, is excited by ultrashort SWIR laser pulses that are tightly focused onto the nanoantenna using a gold-coated off-axis parabolic mirror with a focal length of $f = 33\ \mathrm{mm}$. The illumination is performed in a backside geometry, in which the laser beam passes through the silicon substrate before reaching the nanoantenna located on the front surface.

The photoemitted electrons from the nanoantenna are analyzed using an electron lens system (Focus GmbH), which images the electrons onto a detector plane for either momentum- or real-space microscopy. The lens system consists of an extractor (objective lens), a transfer lens, and two projection lenses (see Extended Data Figure 2). Depending on the voltages applied to the lenses, the operating mode can be switched such that either the initial electron momentum distribution or the electron emission site is imaged onto the detector plane. Accordingly, the same electron-optical system is commonly referred to as either a momentum microscope (MM) or a photoemission electron microscope (PEEM), although both terms describe identical electron optics operated in different imaging modes. In both modes, intermediate momentum- and real-space image planes are formed, allowing physical apertures (contrast and iris apertures) to be inserted. These apertures enable the selection of specific electron trajectories, permitting dark-field imaging by filtering electron momenta or electron micro-spectroscopy by selecting specific emission sites.

At the detector plane, the electron distribution is recorded using a detector assembly consisting of two stacked microchannel plates (MCPs) for electron amplification, a luminescent phosphor screen, and a two-dimensional complementary metal-oxide-semiconductor (CMOS) camera. The camera integration time is set to 10 s, corresponding to an average over $2 \times 10^6$ consecutive laser pulses per image. The brightness of each pixel is a measure of the detected electron count rate. To determine the absolute number of emitted electrons, the integration time was reduced to 1 ms while simultaneously lowering the laser repetition rate to 2 kHz, corresponding to two laser pulses per recorded image. Under these conditions, individual electron impacts could be spatially resolved on the camera, enabling direct counting of detected electrons for the data presented in Figure 2d, even at photoemission yields exceeding 100 electrons per laser pulse. We verified that neither the laser pulse energy nor the pulse duration changed upon reducing the repetition rate. Furthermore, we confirmed that the momentum distributions shown in Figure 2a–c remained unchanged.

**Estimation of local near-field intensities.**

Estimating the local near-field intensity at the antenna apex requires not only the field enhancement provided by the antenna but also knowledge of the peak intensity in the laser focus $I_{\mathrm{focus}}$. Latter is calculated from the measured laser power $P$ in front of the PEEM/MM setup:

$$I_{\mathrm{focus}} = 0.45 \times P/(f\tau A)$$

The laser power is corrected for absorption and reflection losses due to propagation through a replica of the silicon substrate, to account for the incident power at the sample plane. Here, $f = 200\ \mathrm{kHz}$ is the laser repetition rate, $\tau = 20\ \mathrm{fs}$ (FWHM) is the pulse duration and $A$ is the laser spot size in the sample plane.

We define the spot size by $A = \pi(d_{\mathrm{FWHM}}/2)^2$, where $d_{\mathrm{FWHM}}$ is the diameter of the focus given by the FWHM of a crosscut through the intensity profile. The prefactor of 0.45 is a consequence of an assumed Gaussian-pulse profile in space and time (resulting in a factor of 0.65) and considering the relative energy stored in the side pulses from the measured temporal profile (resulting in a factor of 0.69).

The laser spot size is determined directly *in situ* using the imaging capability of the PEEM. For this, the sample is translated laterally such that the laser illuminates only the silicon substrate instead of the gold flake. Unlike gold, silicon is transparent for the SWIR laser pulses. As a result, photoelectrons are emitted from the silicon surface and imaged by the PEEM, revealing a circular photoemission spot that represents the laser focus.

The measured photoemission spot on the silicon substrate, with a diameter of $d_{\mathrm{Si}} = 2.8\ \mu\mathrm{m}$, is smaller than the actual laser focus diameter, $d_{\mathrm{focus}}$, due to the nonlinear photoemission process characterized by the photoemission nonlinearity $n_{\mathrm{Si}}$. The actual focus diameter is therefore obtained from $d_{\mathrm{focus}} = \sqrt{n_{\mathrm{Si}}} d_{\mathrm{Si}}$.

The photoemission nonlinearity of the silicon surface is determined by measuring the total photoelectron yield from the illuminated area as a function of laser power in the multiphoton photoemission regime. Fitting the resulting power dependence yields a nonlinearity of $n_{Si} = 7.4$. Consequently, the laser focus diameter is estimated to $d_{\mathrm{focus}} = 7.7\ \mu\mathrm{m}$.

**Simulations for the nanoantenna field enhancement.**

To simulate the field enhancement factor of the double-nanohole structure, we performed finite-difference time-domain (FDTD) simulations using the commercially available ANSYS Lumerical FDTD software package. The simulated 3D unit cell, with dimensions of 4 × 2 × 0.6 $\mu m^3$, consists of a gold thin film containing two holes, deposited on a bulk silicon substrate. The geometrical parameters were set to match those of the experimental sample, with a hole diameter of 660 nm. The edges of the holes were rounded with a radius of curvature of 9 nm, based on the evaluation of curvature distributions obtained from SEM images. The optical data for Si and Au were taken from the literature [45]. Perfectly matched layer (PML) boundary conditions were applied on all sides of the simulation domain. To accurately describe the near-field distribution around the nanostructure, a mesh size of 1 × 1 × 1 $nm^3$ was used in the vicinity of the apices.

The structure was illuminated with a linearly polarized few-cycle pulse centered at a wavelength of 2000 nm. To determine the field enhancement of the nanostructure, field distributions were collected using frequency-domain field and power monitors positioned at different heights. Since the amplitude of the incident light pulse was set to unity, the resulting field-distribution maps directly represent the field enhancement factor. Field distributions at two different heights are shown in Extended Data Figure 4: at $z = 28\ \mathrm{nm}$, corresponding to the position of the maximum field enhancement, and at $z = 30\ \mathrm{nm}$, corresponding to the top of the structure.

**Three-dimensional electron near-field trajectory simulation.**

The simulation of the in-plane momentum distributions from the nanoantenna is based on the calculation of classical electron trajectories in a near-field within the three-dimensional simple man model (see Figure 3 and Extended Data Figure 5).

The emitting region of one antenna apex is modelled by a metallic nanospheroid and its optical near-fields. The nanospheroid is described by its radii $r_{x,y,z}$ and its surface is parametrized by

$$\frac{x^2}{r_x^2}+\frac{y^2}{r_y^2}+\frac{z^2}{r_z^2}=1.$$

Here, the $x$- and $y$-coordinates describe the sample plane, with $x$ oriented along the antenna direction and $z$ in direction of the electron detector. The nanospheroid model is inspired by inspecting typical ion-milled structures from a tilted view in a scanning electron microscope. The side profile of these structures is well approximated by a rectangular shape with rounded edges, with surface normal vectors that are inclined out of the sample plane. Accordingly, we choose $r_x = r_z$ to accurately approximate the rounded edge by a circle, while $r_y$ is varied to generate different geometries ranging from rounded emitters with a flat surface ($r_y > r_{x,z}$) to strongly curved, tip-like emitters ($r_y < r_{x,z}$).

Because the rounded emitting edge shows out of the sample plane, the locally enhanced optical near-field is expected to point in a similar out-of-plane direction [13]. In practice, the analytical field was calculated for excitation polarized along the $x$-direction and the final electron momenta were subsequently rotated in the $x$-$z$-plane to account for the realistic orientation of the field.

The optical near-field of the nanospheroid is obtained by an analytic solution of the Laplace equation in ellipsoidal coordinates using the quasistatic approximation, yielding the scalar potential [46, 47]. The nanospheroid material is set to gold with an assumed relative permittivity of $\varepsilon_{\mathrm{Au}} = -180$ at 2000 nm wavelength surrounded by vacuum with permittivity $\varepsilon_{\mathrm{Vac}} = 1$, which is excited by an electric field polarized in $x$-direction. The electric near-fields are numerically calculated from the spatial gradients of the scalar potential using finite differences. The calculated fields were normalized such that the maximum amplitude at the nanospheroid surface corresponds to the prescribed local near-field amplitude $E_{\mathrm{NF}}$.

We only consider the spatially localized near-field from the analytical nanospheroid model for the electron propagation. The spatially homogeneous contribution from the incident excitation field is neglected. This approximation is justified by the previously described FDTD simulations of the realistic antenna geometry, predicting field enhancement factors of ~20. Therefore, the incident field without field enhancement is significantly weaker and the electron propagation is dominated by the enhanced near-field.

The total electric near-field from the quasi-static approximation is separated in its spatial and temporal components:

$$\vec{E}(\vec{r},t) = \vec{E}_{\mathrm{NF}}(\vec{r})f(t)$$

The temporal part $f(t)$ is described by a Fourier-limited Gaussian pulse with a pulse duration of $\tau = 20\,\mathrm{fs}$ centered at a wavelength of $\lambda = 2000\,\mathrm{nm}$, $\omega = 2\pi c/\lambda$, and a carrier-envelope phase of $\phi_{\mathrm{CEP}} = \pi$:

$$f(t) = \exp\left(-2\ln 2\frac{t^2}{\tau^2}\right)\cos(\omega t + \phi_{\mathrm{CEP}})$$

Electrons are initially placed at different positions on the nanospheroid surface by sampling the azimuthal angle $\varphi$ and polar angle $\theta$. The emission is restricted to relevant regions around the apex, by limiting both angular coordinates. The initial sampling spans from $10° \leq \varphi \leq 170°$ and $60° \leq \theta \leq 120°$, in steps of 3°. Additionally, only angle combinations that are located within an elliptical region centered at $\varphi, \theta = 90°$ were considered, such that they fulfil

$$\left(\frac{\theta - 90°}{\Delta\theta}\right)^2 + \left(\frac{\varphi - 90°}{\Delta\varphi}\right)^2 < 1.$$

Here, $\Delta\theta = 30°$ and $\Delta\varphi = 80°$ are the angular half-widths of the sampled distributions. This constraint excludes positions on the rear and peripheral parts of the nanospheroid that would lie inside the bulk antenna. This effectively limits the initial electron positions to the rounded apex region which is intended to represent the experimentally emitting antenna edge.

The angular coordinates are converted to positions on the surface according to

$$x = r_x \sin\theta \cos\varphi\,, \quad y = r_y \sin\theta \sin\varphi, \quad z = r_z \cos\theta,$$

and the electrons are initialized with a starting velocity of zero.

Initial emission times were sampled in steps of 0.02 fs over the entire temporal window. For each combination of emission position and time, the instantaneous local electric field was calculated.

Each electron was assigned a field-dependent emission probability obtained from the analytic Keldysh rate

$$P_{\mathrm{field}} \propto \exp\left[-\frac{2\Phi}{\hbar\omega}\left(\left[1+\frac{1}{2\gamma^2}\right]\operatorname{asinh}(\gamma) - \frac{\sqrt{1+\gamma^2}}{2\gamma}\right)\right],$$

with the local Keldysh parameter

$$\gamma(\vec{r},t) = \frac{\omega\sqrt{2m\Phi}}{\left|q\vec{E}(\vec{r},t)\right|}.$$

Here, we set the work function of gold to $\Phi = 3.5\ \mathrm{eV}$, which is obtained from the ratio of the measured nonlinearity in the regime of multiphoton photoemission (see Figure 2d) and the photon energy. The physical constants $m$ and $q$ denote the electron mass and charge, respectively. Emission was only permitted during optical half-cycles for which the electric force acting on the electrons was directed away from the surface. Initial conditions with a relative probability below $10^{-2}$ of the maximum emission probability were omitted to reduce the number of propagated trajectories.

Each starting position is weighted by the local field-dependent emission probability and by the surface-area element

$$P_{\mathrm{surface}} \propto r_x \sin\theta \sqrt{r_x^2 \cos^2\varphi \sin^2\theta + r_y^2 \sin^2\varphi \sin^2\theta + r_y^2 \cos^2\theta}.$$

Weighting by the surface area compensates for the non-uniform surface area that results from a uniform spacing in polar and azimuthal coordinates.

After emission, each electron is propagated independently in the near-field as a classical particle by solving the equation of motion

$$m\frac{\mathrm{d}^2\vec{r}(t)}{\mathrm{d}t^2} = q\vec{E}(\vec{r},t),$$

using an adaptive ordinary-differential equation solver.

Recollisions with the surface were modeled as elastic scattering events that conserve the electron's kinetic energy. The outgoing direction was randomized within a finite angular range, while ensuring that the electron was scattered away from the surface, to account for the finite angular scattering distribution and atomic-scale surface roughness.

The calculated electron trajectories are classified based on their distance from the nanospheroid surface as a function of time. Trajectories without a reversal of their outward motion are classified as subcycle electrons, whereas trajectories with a reversal are classified as quiver electrons. Electrons for which a collision with the surface was detected are classified as rescattered electrons.

After the optical pulse has vanished, the final electron momenta $\vec{k}$ are calculated from the final velocities $\vec{v}$ using the electron mass $m$ and the reduced Planck constant $\hbar$

$$\vec{k} = \frac{m\vec{v}}{\hbar}$$

and are rotated by 50° in the $xz$-plane to account for the orientation of the field. Only electrons with a positive final momentum in detector direction ($z$) were retained.

Finally, simulated in-plane momentum distributions were obtained from weighted two-dimensional histograms of the final $k_x$ and $k_y$ components with weights $\propto P_{\mathrm{field}}P_{\mathrm{surface}}$. To model the complete nanoantenna, the distribution calculated for one apex is mirrored along the apex direction and added to the original distribution. The resulting momentum maps were smoothed with a Gaussian kernel to remove graininess from the finite sampling of initial electron conditions.

**Extraction of the width of the energy distribution from in-plane momenta.**

We extract the width of the energy distribution from a cross section through the center of the measured and simulated in-plane momentum distributions at $k_y = 0$. The positions $k_{x+}$ and $k_{x-}$ at which the resulting line profile reaches 50% of its maximum intensity are determined on the positive and negative $k_x$ sides, respectively. These positions correspond to the two half-maximum positions of the momentum distribution, and we expect the electrons with the highest energy to contribute predominantly to this cross section because they originate near the center of the antenna apex, where the local near-field amplitude is the largest. Electrons emitted farther away from the apex center experience weaker fields and generally acquire lower kinetic energies and larger transverse momentum components.

Because the emitting antenna edge is assumed to be inclined by ~50° with respect to the sample plane, we assume equal momentum components in apex direction and in detector direction $k_x \approx k_z$. The half-maximum points are then converted to energies via

$$E_{kin} = \frac{\hbar^2}{2m}(k_x^2 + k_z^2) \approx \frac{\hbar^2}{m} k_x^2$$

for both $k_x$-directions, yielding the widths of the energy distribution.

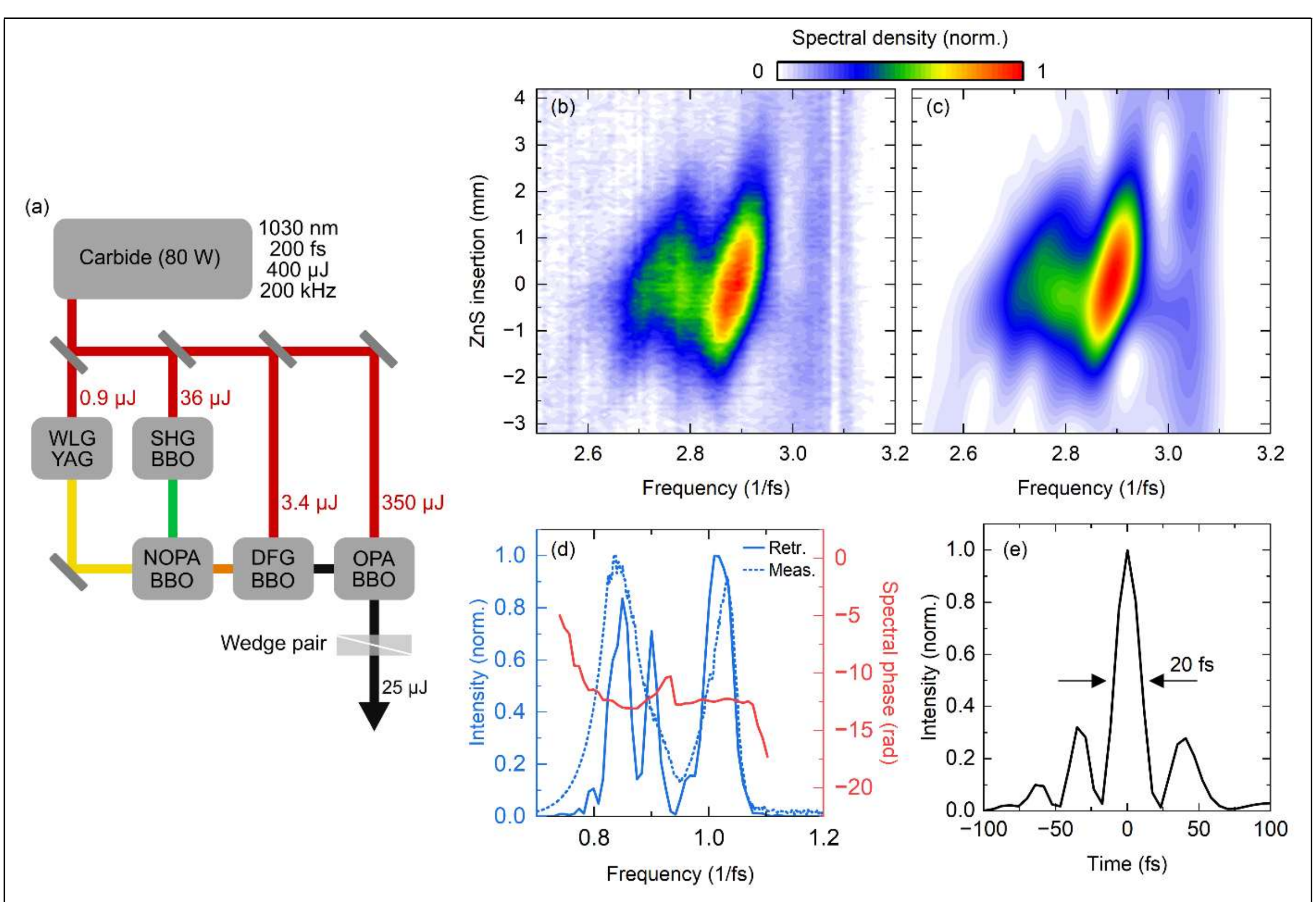

**Extended Data Figure 1:** Laser setup and characterization. **(a)** Sketch of the amplifier system. The fundamental laser beam is divided into four separate parts, used for pumping different nonlinear conversion stages. We generate few-cycle laser pulses in the short-wavelength infrared with a pulse energy of 22.5 μJ centered around a wavelength of 2000 nm. **(b)** Measured dispersion scan (d-scan) spectrogram for pulse characterization, obtained by measuring the third harmonic spectrum of the beam generated at a glass-air interface while simultaneously inserting a dispersive zinc sulfide (ZnS) wedge pair. **(c)** Simulated spectrogram using a retrieval algorithm. **(d)** Measured (dotted blue line) and retrieved (solid blue line) spectrum of the laser pulses and retrieved spectral phase (solid red line). **(e)** Retrieved temporal intensity profile of the pulses, showing a pulse duration of 20 fs (FWHM).

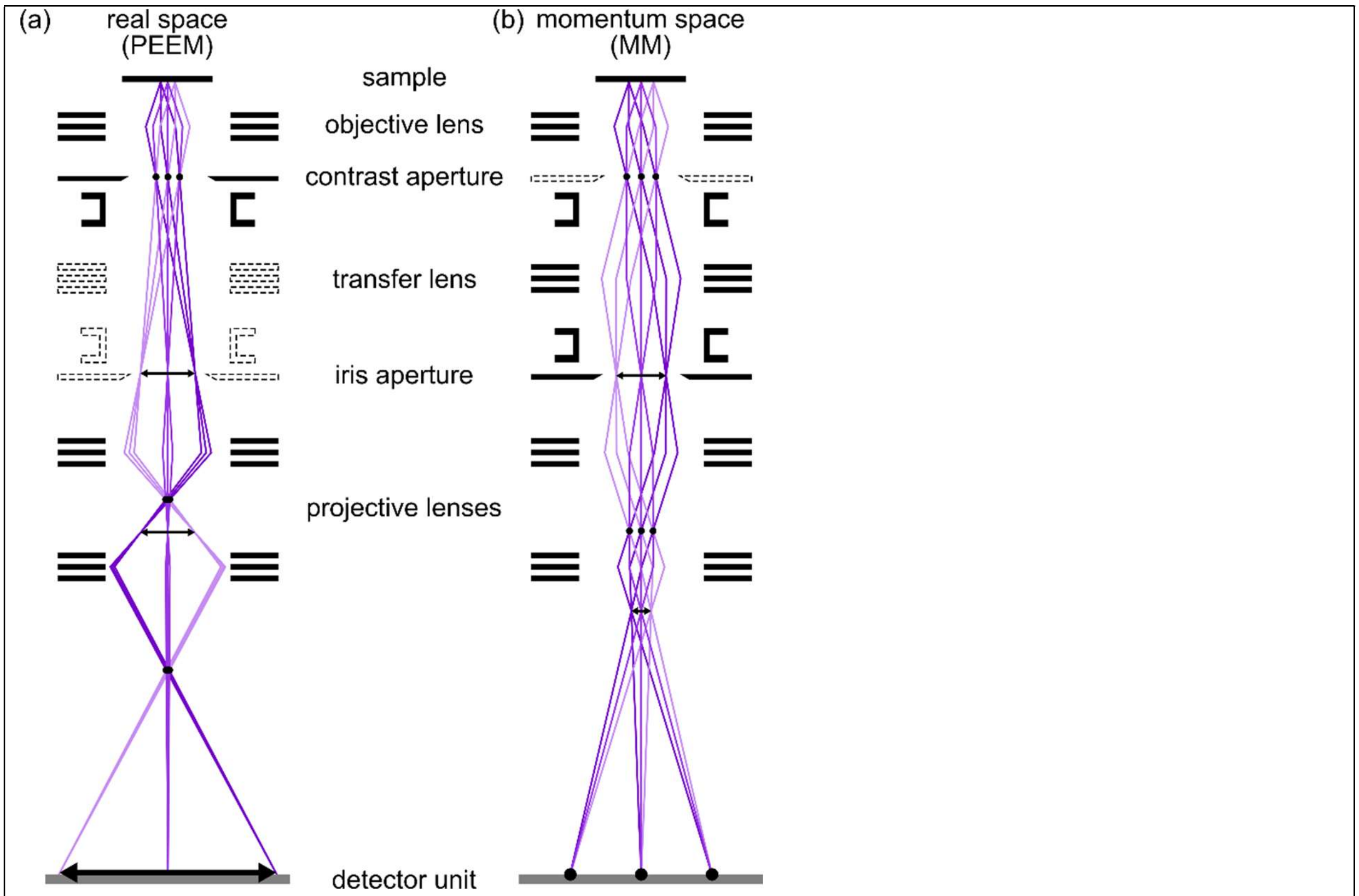

**Extended Data Figure 2:** Electron-optical path through the lens columns. **(a, b)** Electron path through the column for photoemission electron microscopy (PEEM, Focus GmbH), enabling real-space imaging (a) and for momentum microscopy (MM) allowing for momentum-space imaging (b). In PEEM-mode a contrast aperture can be inserted to filter specific electron momenta, while an iris aperture can be inserted in MM-mode to filter emission positions.

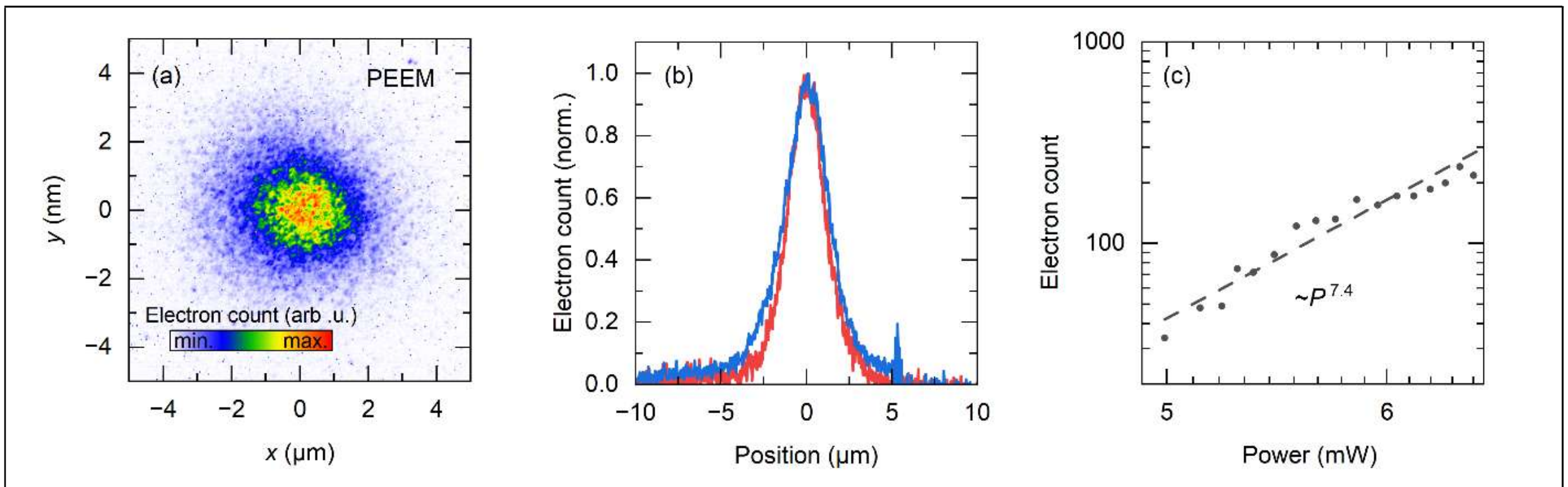

**Extended Data Figure 3:** Determination of laser focus size. **(a)** Measured photoemission spot by focusing the laser pulses on the silicon substrate and triggering nonlinear multiphoton photoemission. **(b)** Spot profiles by integrating the photoemission yield from (a) in vertical (blue) and horizontal (red) direction, revealing a Gaussian-like profile with approximately equal diameter in both directions. **(c)** Measured nonlinear photoemission yield from the silicon surface. A linear fit reveals a nonlinearity of $n = 7.4$, which is used to convert the photoemission spot to a focus size.

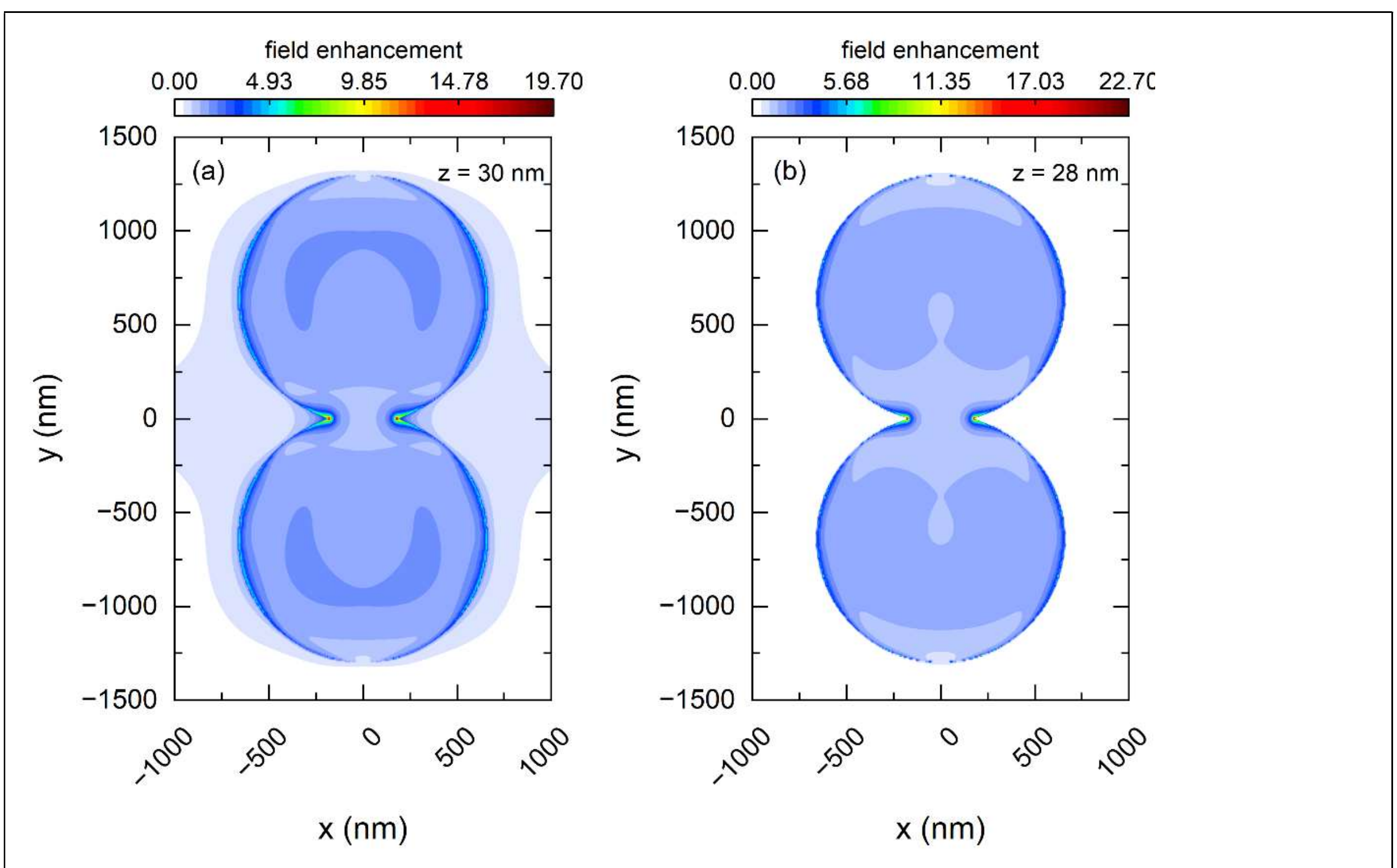

**Extended Data Figure 4:** Numerical determination of the field enhancement. **(a, b)** Field distribution maps collected at $z = 30\,\mathrm{nm}$ (a), corresponding to the top of the structure, and at $z = 28\,\mathrm{nm}$ (b), corresponding to the top of the position of the maximum field enhancement.

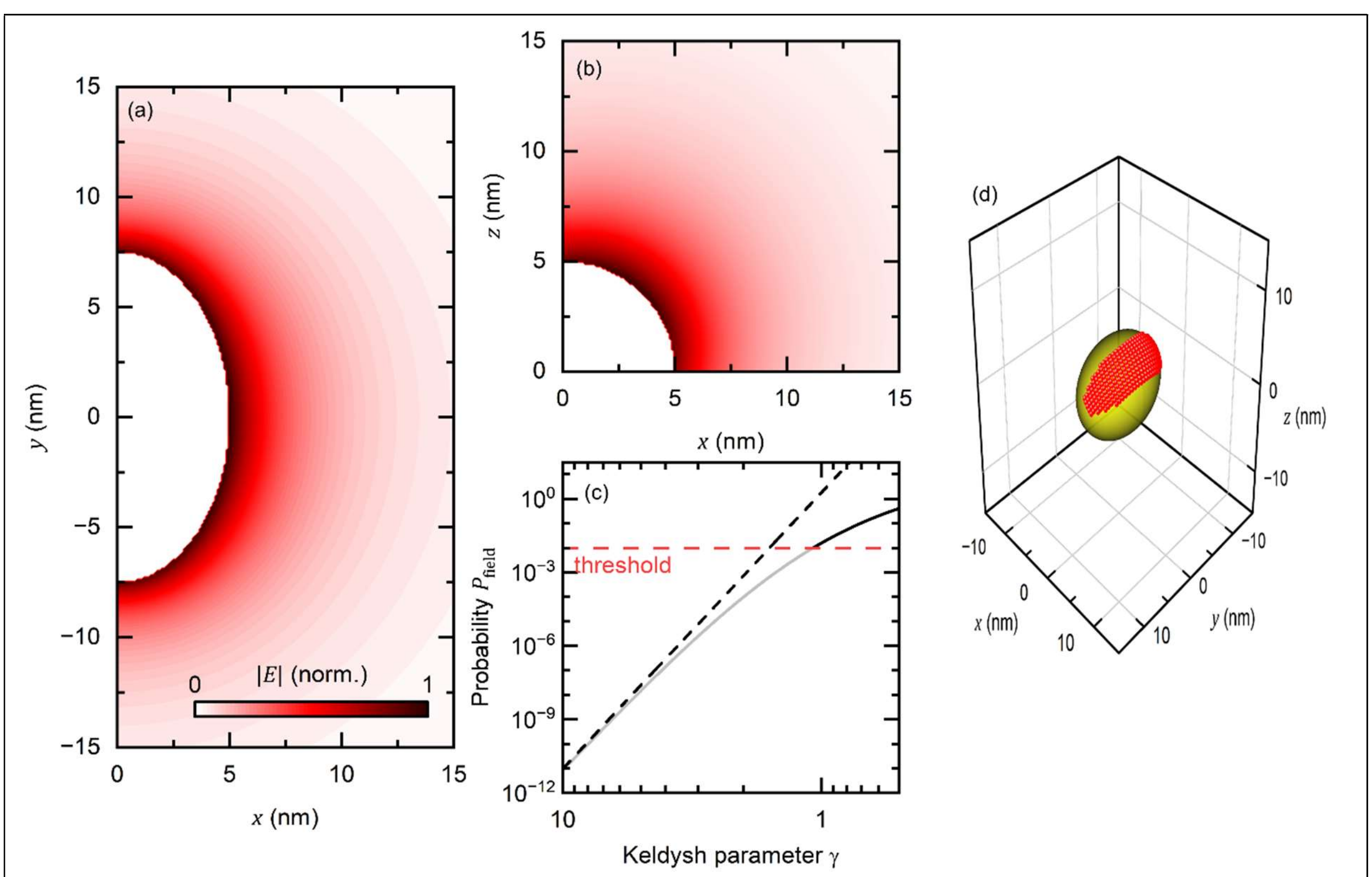

**Extended Data Figure 5:** Nanospheroid model for electron trajectories. **(a, b)** Optical near-field around the nanospheroid in the $x$-$y$-plane (a) and $x$-$z$-plane (b). **(c)** Photoemission probability from the Keldysh model (solid line) and a multiphoton photoemission rate with nonlinearity of 5.6 (black dashed line). Photoelectrons with a photoemission probability below the threshold (red dashed line) are not considered in the simulations. **(d)** Initial electron positions on the nanospheroid surface.

## References

1. Choi, D., et al., *Observation of Floquet–Bloch states in monolayer graphene.* Nature Physics, 2025. **21**(7): p. 1100–1105.
2. Merboldt, M., et al., *Observation of Floquet states in graphene.* Nature Physics, 2025. **21**(7): p. 1093–1099.
3. Wang, Y.H., et al., *Observation of Floquet-Bloch States on the Surface of a Topological Insulator.* Science, 2013. **342**(6157): p. 453–457.
4. Lesko, D.M.B., et al., *Probing Broken Time-Reversal Symmetry in 2D Materials with Tailored-Light Photocurrent Generation.* ACS Nano, 2026. **20**(15): p. 11614–11623.
5. Mahmood, F., et al., *Selective scattering between Floquet–Bloch and Volkov states in a topological insulator.* Nature Physics, 2016. **12**(4): p. 306–310.
6. Dombi, P., et al., *Strong-field nano-optics.* Reviews of Modern Physics, 2020. **92**(2): p. 025003.

7. Ciappina, M.F., et al., *Attosecond physics at the nanoscale.* Reports on Progress in Physics, 2017. **80**(5): p. 054401.
8. Krüger, M., M. Schenk, and P. Hommelhoff, *Attosecond control of electrons emitted from a nanoscale metal tip.* Nature, 2011. **475**(7354): p. 78–81.
9. Herink, G., et al., *Field-driven photoemission from nanostructures quenches the quiver motion.* Nature, 2012. **483**(7388): p. 190–193.
10. Park, D.J., et al., *Strong Field Acceleration and Steering of Ultrafast Electron Pulses from a Sharp Metallic Nanotip.* Physical Review Letters, 2012. **109**(24): p. 244803.
11. Spektor, G., et al., *Revealing the subfemtosecond dynamics of orbital angular momentum in nanoplasmonic vortices.* Science, 2017. **355**(6330): p. 1187–1191.
12. Großmann, M., et al., *Light-Triggered Control of Plasmonic Refraction and Group Delay by Photochromic Molecular Switches.* ACS Photonics, 2015. **2**(9): p. 1327–1332.
13. Rácz, P., et al., *Measurement of Nanoplasmonic Field Enhancement with Ultrafast Photoemission.* Nano Letters, 2017. **17**(2): p. 1181–1186.
14. Schmidt, O., et al., *Time-resolved two photon photoemission electron microscopy.* Applied Physics B, 2002. **74**(3): p. 223–227.
15. Paschen, T., et al., *Ultrafast Strong-Field Electron Emission and Collective Effects at a One-Dimensional Nanostructure.* ACS Photonics, 2023. **10**(2): p. 447–455.
16. Budai, J., et al., *Plasmon–plasmon coupling probed by ultrafast, strong-field photoemission with <7 Å sensitivity.* Nanoscale, 2018. **10**(34): p. 16261–16267.
17. Chen, K., et al., *High-Q, low-mode-volume and multiresonant plasmonic nanoslit cavities fabricated by helium ion milling.* Nanoscale, 2018. **10**(36): p. 17148–17155.
18. Schurr, B., et al., *Plasmonic Su–Schrieffer–Heeger chains with strong coupling amplitudes.* Science Advances, 2025. **11**(50): p. eaea3844.
19. Ropers, C., et al., *Localized Multiphoton Emission of Femtosecond Electron Pulses from Metal Nanotips.* Physical Review Letters, 2007. **98**(4): p. 043907.
20. Hommelhoff, P., et al., *Field Emission Tip as a Nanometer Source of Free Electron Femtosecond Pulses.* Physical Review Letters, 2006. **96**(7): p. 077401.
21. Hergert, G., et al., *Ultra-Nonlinear Subcycle Photoemission of Few-Electron States from Sharp Gold Nanotapers.* Nano Letters, 2024. **24**(35): p. 11067–11074.
22. Hergert, G., R. Lampe, and C. Lienau, *Quenching Strong-Field Rescattering of Photoemitted Electrons from Metallic Nanotapers by Using Moderate Bias Fields.* ACS Photonics, 2025. **12**(4): p. 2219–2225.
23. Hergert, G., et al., *Probing Transient Localized Electromagnetic Fields Using Low-Energy Point-Projection Electron Microscopy.* ACS Photonics, 2021. **8**(9): p. 2573–2580.
24. Zhong, J.-H., et al., *Nonlinear plasmon-exciton coupling enhances sum-frequency generation from a hybrid metal/semiconductor nanostructure.* Nature Communications, 2020. **11**(1): p. 1464.
25. Hergert, G., et al., *Long-lived electron emission reveals localized plasmon modes in disordered nanosponge antennas.* Light: Science & Applications, 2017. **6**(10): p. e17075–e17075.

26. Komatsu, K., et al., *Few-Cycle Surface Plasmon Polaritons.* Nano Letters, 2024. **24**(8): p. 2637–2642.
27. Schmitt, D., et al., *Ultrafast nano-imaging of dark excitons.* Nature Photonics, 2025. **19**(2): p. 187–194.
28. Brückner, L., et al., *A Gold Needle Tip Array Ultrafast Electron Source with High Beam Quality.* Nano Letters, 2024. **24**(16): p. 5018–5023.
29. Dombi, P., et al., *Ultrafast Strong-Field Photoemission from Plasmonic Nanoparticles.* Nano Letters, 2013. **13**(2): p. 674–678.
30. Teichmann, S.M., et al., *Strong-field plasmonic photoemission in the mid-IR at <1 GW/cm2 intensity.* Scientific Reports, 2015. **5**(1): p. 7584.
31. Schötz, J., et al., *Onset of charge interaction in strong-field photoemission from nanometric needle tips.* Nanophotonics, 2021. **10**(14): p. 3769–3775.
32. Bormann, R., et al., *Tip-Enhanced Strong-Field Photoemission.* Physical Review Letters, 2010. **105**(14): p. 147601.
33. Boroviks, S., et al. *Crystalline metal flakes: Platforms for advanced plasmonics and hybrid 2D material architectures*. 2026. arXiv:2604.22988 DOI: 10.48550/arXiv.2604.22988.
34. Meier, K., et al., *Multiscale carrier-envelope phase characterization of 2-µm pulses delivered by a 200-kHz optical parametric amplifier.* Applied Physics B, 2026. **132**(8): p. 97.
35. Anderson, P.A., *Work Function of Gold.* Physical Review, 1959. **115**(3): p. 553–554.
36. Kollmann, H., et al., *Toward Plasmonics with Nanometer Precision: Nonlinear Optics of Helium-Ion Milled Gold Nanoantennas.* Nano Letters, 2014. **14**(8): p. 4778–4784.
37. Bánhegyi, B., et al., *Nanoplasmonic Photoelectron Rescattering in the Multiphoton-Induced Emission Regime.* Physical Review Letters, 2024. **133**(3): p. 033801.
38. Paulus, G.G., et al., *Plateau in above threshold ionization spectra.* Physical Review Letters, 1994. **72**(18): p. 2851–2854.
39. Cornaggia, C., *Molecular rescattering signature in above-threshold ionization.* Physical Review A, 2008. **78**(4): p. 041401.
40. Park, D.J., et al., *Characterizing the optical near-field in the vicinity of a sharp metallic nanoprobe by angle-resolved electron kinetic energy spectroscopy.* Annalen der Physik, 2013. **525**(1-2): p. 135–142.
41. Echternkamp, K.E., et al., *Strong-field photoemission in nanotip near-fields: from quiver to sub-cycle electron dynamics.* Applied Physics B, 2016. **122**(4): p. 80.
42. Miranda, M., et al., *Simultaneous compression and characterization of ultrashort laser pulses using chirped mirrors and glass wedges.* Optics Express, 2012. **20**(1): p. 688–697.
43. Sytcevich, I., et al., *Characterizing ultrashort laser pulses with second harmonic dispersion scans.* Journal of the Optical Society of America B, 2021. **38**(5): p. 1546–1555.
44. Geib, N.C., et al., *Common pulse retrieval algorithm: a fast and universal method to retrieve ultrashort pulses.* Optica, 2019. **6**(4): p. 495–505.
45. Haynes, W.M., *CRC handbook of chemistry and physics*. 2016: CRC press.

46. Bohren, C.F. and D.R. Huffman, *Absorption and scattering of light by small particles*. 2008: John Wiley & Sons.
47. Landau, L.D., et al., *Electrodynamics of continuous media*. 1961, American Institute of Physics.